\documentstyle[epsfig]{elsart}
\def\deca{Z \to \nu \bar{\nu} \gamma}
\def\decb{Z \to \nu \bar{\nu} \gamma \gamma}

\begin{document}
\begin{frontmatter}
\title{Constraints on neutrino-photon interactions from rare $Z$ decays
}
\author[CINVESM]{F. Larios \thanksref{emailfl},}
\author[CINVES]{M. A. P\'erez \thanksref{emailmap},} and
\author[CINVES]{G. Tavares-Velasco \thanksref{emailgtv}}
\address[CINVESM]{Departamento de F\'\i sica Aplicada, CINVESTAV-M\' erida,
A. P. 73, 91310, M\' erida, Yucat\' an, M\' exico.}
\address[CINVES]{Departamento de F\'{\i}sica, CINVESTAV
IPN, Apartado Postal 14-740, 07000, M\'exico D. F., M\'exico.}

\thanks[emailfl]{E-address: flarios@belinda.cinvestav.mda.mx}
\thanks[emailmap]{E--address: mperez@fis.cinvestav.mx}
\thanks[emailgtv]{E--address: gtv@fis.cinvestav.mx}

\begin{abstract}

It is shown that the rare decays $\deca$ and $\decb$ are useful to
put model-independent bounds on neutrino-one-photon and
neutrino-two-photon interactions. The results are then used to
constrain the $\tau$ neutrino magnetic moment $\mu_{\nu_\tau}$ and
the double radiative decay $\nu_j\to\nu_i\gamma \gamma$. It is
found that the decay $\deca$ gives a more stringent bound on
$\mu_{\nu_\tau}$ than that obtained from $\decb$; the latter decay
in turn gives limits on the neutrino-two-photon interaction that
are less stringent than those obtained for a sterile neutrino
$\nu_s$ from the analysis of $\nu_\mu N\to \nu_s N$ conversion.

\end{abstract}

\end{frontmatter}

\maketitle

The behavior recently observed of atmospheric \cite{Fukuda} and
solar \cite{Fukuda1} neutrinos provides rather strong evidence
that neutrinos have mass. This fact has renewed the interest in
neutrino electromagnetic properties, which have received
considerable attention as they may shed light on some physics
issues. In parti\-cular, neutrino-one-photon interactions are of
interest since they may play a key role in elucidating the solar
neutrino puzzle, which can be explained in part by a large
neutrino magnetic moment \cite{Cisneros}. In the simplest
extension of the standard model (SM), with the presence of massive
neutrinos, one-loop radiative corrections induce a small magnetic
moment proportional to the neutrino mass $m_\nu$, i.e.
$\mu_\nu=3\,e\,G_F\,m_\nu/(8\,\sqrt{2}\,\pi^2)=3\times 10^{-19}\,
m_\nu\; \mu_{\mathrm B}$ \cite{Fujikawa}, where $m_\nu$ is to be
expressed in KeV and $\mu_{\mathrm B}$ stands for the Bohr
magneton. Several models  have been advanced in order to induce
neutrino magnetic moments as large as $10^{-11}$-$10^{-10}\;\,
\mu_{\mathrm B}$ \cite{Frank}, even with neutrino masses
compatible with the mass square differences needed by atmospheric
\cite{Fukuda}, solar \cite{Fukuda1} and the liquid scintillation
neutrino detector (LSND) \cite{Athana} data. As for
neutrino-two-photon interactions, they may have direct
implications on several low- and high-energy reactions with
astrophysical and cosmological interest \cite{Pontecorvo}. For
instance, a high annihilation rate of photons into a neutrino pair
may explain the observed cooling of stars by neutrino emission
\cite{Levine}. In addition, there are other interesting processes
involving neutrino-two-photon interactions, such as $\nu \gamma
\to \nu \gamma$, $\nu \bar \nu \to \gamma \gamma$, and the
neutrino double-radiative decay $\nu_j\to \nu_i \gamma \gamma$. It
is important to note that in the SM with massive neutrinos, the
decay $\nu_j\to \nu_i \gamma \gamma$ is not severely suppressed by
the GIM mechanism and can be the main decay channel as long as the
$\nu_j$ mass lies in the range of a few tenths of a MeV
\cite{Nieves}.

From the experimental side, the L3 collaboration searched for
single-photon events near the $Z$ pole at the CERN LEP collider
and set a bound on the rare decay $\deca$ \cite{Acciari}. It was
shown that the collected data impose a stringent constraint on the
$\tau$ neutrino magnetic moment \cite{Acciari,Maya,Maltoni}. In
fact, the decay $\deca$ can be a valuable tool to search for
evidences of new physic since its rate is negligibly small in the
SM \cite{Hernandez}. By using the experimental bound on $\deca$,
an analysis in the framework of the effective Lagrangian approach
(ELA) was carried out in Refs. \cite{Maya,Larios} in order to
constrain the operators that induce the couplings $\nu \bar \nu
\gamma$, $\nu \bar \nu Z \gamma$ and $ZZ\gamma$. In regard to the
rare decay $\decb$, long ago the L3 and the OPAL collaborations
looked for events with a lepton pair accompanied by a photon pair
of large invariant mass \cite{OPAL}. After combining the data of
both searches, the OPAL collaboration set an upper bound on the
rate of $\decb$: it was found that BR$(\decb)\leq3.1 \times
10^{-6}$.

In the present letter we consider the possibility of obtaining
indirect bounds on neutrino electromagnetic interactions from the
experimental constraints on the decays $\deca$ and $\decb$. Our
main goal is to study these processes in a model independent way.
We will also show that our results can be used to constrain the
$\nu_\tau$ magnetic moment and the decay $\nu_j \to \nu_i
\gamma \gamma$.

For the purpose of this letter we will consider the following
effective interaction

\begin{equation}
{\cal L}_{\bar \nu_i  \nu_j \gamma}=\frac{1}{2}\, \mu_{\nu_i
\nu_j} \,\bar \nu_i\, \sigma_{\mu \nu} \,\nu_j F^{\mu\nu},
\label{Lnng}
\end{equation}

\noindent where $\mu_{\nu_i}\equiv\mu_{\nu_i\nu_i}$ is the $\nu_i$
magnetic moment and $\mu_{\nu_i \nu_j}$ ($i\ne j$) is the
transition magnetic moment. Although we will focus on Dirac
neutrinos here, the discussion can be readily extended to
Majorana neutrinos.

\begin{figure}
\begin{center}
\epsfig{file=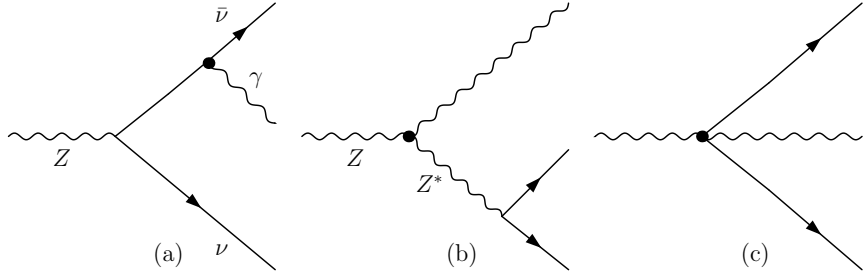,width=4.5in} \caption{Feynman diagrams
contributing to the decay $Z \to \nu \overline{\nu} \gamma$ in the
effective Lagrangian  approach. The dots denote effective
couplings.} \label{znng}
\end{center}
\end{figure}

As already mentioned, in Ref. \cite{Hernandez} it was shown that
the SM rate of the decay $\deca$ is unobservably small. Therefore,
it represents an extraordinary mode to look for evidences of new
physics arising from neutrino-one-photon interactions at a future
$e^+e^-$ linear collider. The search for events with ener\-getic
single-photons along with missing energy at LEP was used by the L3
collaboration to set the bound BR$(\deca)\leq 10^{-6}$
\cite{Acciari}. Within the ELA, the rare decay $\deca$ can proceed
through the Feynman diagrams shown in Fig. \ref{znng}. For details
of the analysis of these diagrams, we refer the reader to Refs.
\cite{Maya,Larios}. In particular, the experimental limit on
$\deca$ gives the following bound on the $\nu_\tau$ magnetic
moment

\begin{equation}
\mu_{\nu_\tau}\le 2.62 \times 10^{-6} \;\mathrm{\mu_B}.
\label{bound1}
\end{equation}

\noindent This bound is in good agreement with that found by the
L3 collaboration \cite{Acciari}, and compares favorably with the
bounds $\mu_{\nu_{\tau}}< \mathrm{4} \times \mathrm{10^{-6}}
\,\;\mathrm{\mu_B}$ \cite{Grotch} and $\mu_{\nu_{\tau}}<
\mathrm{2.7} \times \mathrm{10^{-6}} \,\;\mathrm{\mu_B}$
\cite{Escribano}. The former was obtained from low-energy
experiments, whereas the latter was derived from the invisible
width of the $Z$ boson. Furthermore, our bound is close to the one
obtained from a beam-dump experiment \cite{Cooper-Sarkar}. It is
important to note that the most stringent bounds on the neutrino
magnetic moment are obtained from chirality flip in supernova
\cite{Ayala}.

We now turn to examine the rare decay $\decb$, which also can
receive contributions from a neutrino-one-photon interaction
through the Feynman diagrams depicted in Fig. 2. We would like to
analyze if this decay is useful to bound the neutrino magnetic
moment. Given the effective interaction of Eq. (\ref{Lnng}), the
amplitude for the Feynman diagrams of Fig. 2 plus the crossed ones
reads

\begin{equation}
{\cal M}={\cal M}^{\alpha\beta\mu}\, \epsilon^*_\alpha(k_1)
\epsilon^*_\beta(k_2) \epsilon_\mu(p)+\ldots \label{MZnngg1}
\end{equation}

\noindent where the ellipsis stands for the crossed diagrams
contribution, which can be obtained from the first term after the
substitutions $\alpha \leftrightarrow \beta$ and
$k_1\leftrightarrow k_2$. In principle, we must take into account
all the neutrino species. However, in order to get an upper bound
on $\mu_{\nu_\tau}$, we will make a few assumptions for the sake
of simplicity. First of all, we consider that the neutrino
magnetic moment matrix is almost flavor diagonal, i.e.
$\mu_{\nu_i}\gg \mu_{\nu_i\nu_j}$ ($j\ne i$). Secondly, we assume
that the $\nu_\tau$ magnetic moment dominates over $\mu_{\nu_e}$
and $\mu_{\nu_\mu}$, i.e. there is the hierarchy
$\mu_{\nu_\tau}\gg \mu_{\nu_\mu} \gg \mu_{\nu_e}$. In fact, the
most stringent experimental bounds are $\mu_{\nu_e}\le 1.1\times
10^{-10}\; \mu_{\mathrm B}$ \cite{Krakauer}, $\mu_{\nu_\mu}\le
7.4\times 10^{-9}\; \mu_{\mathrm B}$ \cite{Krakauer}, and
$\mu_{\nu_\tau}\le 5.4\times 10^{-7}\; \mu_{\mathrm B}$
\cite{Cooper-Sarkar}. Bearing in mind these assumptions, the main
contribution to the $\decb$ rate will arise from the
$\bar{\nu}_\tau \nu_\tau \gamma$ vertex. Therefore we can write
${\cal M}^{\alpha\beta\mu}$ as

\begin{equation}
{\cal M}^{\alpha\beta\mu}=
\frac{i\,g\,\mu_{\nu_\tau}^2}{2\,c_W}\sum_{k=1}^3\bar{\nu}_\tau(p_2)
\Gamma^{\alpha\beta\mu}_k \nu_\tau(p_1)\,{k_1}_\lambda {k_2}_\rho,
\label{MZnngg2}
\end{equation}

\noindent with

\begin{equation}
\Gamma^{\alpha\beta\mu}_1=\gamma^\mu\,P_L\,\sigma^{\alpha\lambda}
\left(\FMSlash{p}-\FMSlash{p}_2\right)^{-1}
\sigma^{\beta\rho}\left(\FMSlash{p}_1+\FMSlash{k}_2\right)^{-1},
\end{equation}

\begin{equation}
\Gamma^{\alpha\beta\mu}_2=\sigma^{\alpha\lambda}\left(\FMSlash{p}_2
+\FMSlash{k}_1\right)^{-1}
\sigma^{\beta\rho}\left(\FMSlash{p}-\FMSlash{p}_1\right)^{-1}\gamma^\mu\,
P_L,
\end{equation}

\begin{equation}
\Gamma^{\alpha\beta\mu}_3=\sigma^{\alpha\lambda}\left(\FMSlash{p}_2
+\FMSlash{k}_1\right)^{-1}
\gamma^\mu\,
P_L\,\sigma^{\beta\rho}\left(\FMSlash{p}_1+\FMSlash{k}_2\right)^{-1},
\end{equation}

\noindent where we have neglected the $\nu_\tau$ mass; $p_1
\,(p_2)$, $k_{1,2}$ and $p$ are the neutrino (antineutrino),
photon and  $Z$ boson four-momenta; and $P_{L}=(1-\gamma^5)/2$ is
the left-handed helicity projector.

The transition amplitude can be squared by the usual trace
technique. The result is too lengthy to be shown here. The squared
amplitude can then be integrated over the four-body phase space
with the aid of the Monte Carlo integration method \cite{VEGAS}.
In order to cross-check our results,  we used two different
methods for the evaluation of the $\decb$ decay rate. In the first
method we squared the amplitude and then used a Monte Carlo event
generator to carry out the numerical integration
\cite{MonteCarlo}. As far as the second method is concerned, we
implemented the $\nu\bar\nu\gamma$ and $\nu\bar\nu\gamma\gamma$
interactions into the {\small CALCHEP} program \cite{COMPHEP},
which automatically generates the respective set of Feynman
diagrams, squares the matrix elements, and integrates over the
phase space. There was nice agreement between the results obtained
by both of these methods.

Under the assumptions discussed above, we can obtain the following
estimate for the $\decb$ rate

\begin{equation}
\mathrm{BR}(Z\to \nu \bar \nu\gamma \gamma)= 1.749 \times
10^{11}\, \left(\frac{\mu_{\nu_\tau}}{1\,\mu_{\mathrm
B}}\right)^4,
\end{equation}

\noindent which along with the experimental bound
BR$(\decb)\leq3.1 \times 10^{-6}$ yield

\begin{equation}
\mu_{\nu_\tau}\leq6.488\times 10^{-5} \;\mu_{\mathrm{B}}.
\label{bound2}
\end{equation}

\noindent which is just one order of magnitud below than the bound
obtained from the three body decay $\deca$ [cf. Eq.
(\ref{bound1})]. The rare decay $\decb$ is however more sensitive
to the value of the neutrino magnetic moment. In this respect, it
is interesting if we take a different approach and use the most
stringent experimental bound on $\mu_{\nu_\tau}$
\cite{Cooper-Sarkar} to constrain the rare decays $\deca$ and
$\decb$, in which case we are led to

\begin{equation}
{\mathrm BR}(\deca)\le 6.917 \times 10^{-8},
\end{equation}

\begin{equation}
{\mathrm BR}(\decb)\le 1.487 \times 10^{-14}.
\end{equation}

\begin{figure}
\begin{center}
\epsfig{file=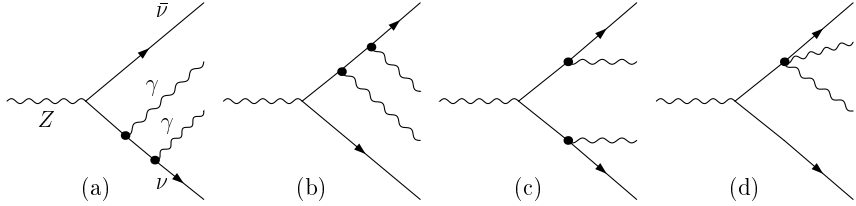,width=4.5in} \caption{Feynman diagrams
contributing to the decay $Z \to \nu \overline{\nu} \gamma
\gamma$. The crossed diagrams are not shown. The dots denote
effective couplings.} \label{znngg}
\end{center}
\end{figure}

Before proceeding, we would like to stress that the procedure
described above can be employed to get bounds on the $\tau$
neutrino transition magnetic moments $\mu_{\nu_\tau \nu_i}$
($i=e,\,\mu$), which also have been the source of interest
\cite{Martinez}. In that case, we would have to include all the
$\mu_{\nu_\tau \nu_i}$ contributions into the $\deca$ and $\decb$
rates, and the upper bound on each $\mu_{\nu_\tau \nu_i}$ would be
obtained after dropping the remaining contributions. Since the
results are insensitive to the neutrino mass, provided that $m_\nu
\ll m_Z$, it follows that the bounds of Eqs. (\ref{bound1}) and
(\ref{bound2}) also apply to $\mu_{\nu_\tau \nu_i}$. Of course the
same is true for any $\mu_{\nu_i \nu_j}$, but the bounds are very
weak for $\nu_e$ and $\nu_\mu$, as compared to other results
appearing in the literature.

Now we will analyze the impact of the $\nu\bar\nu\gamma\gamma$
coupling on the rare decay $\decb$. While the neutrino-one-photon
interaction is strictly vanishing for massless neutrinos, the
neutrino-two-photon interaction can have a nonzero value even if
neutrinos are massless. Long ago, it was shown that if massless
neutrinos interact locally with charged leptons the
neutrino-two-photon vertex vanishes \cite{GellMann}. It is
possible that neutrinos are kept massless but interact directly
with gauge bosons, as in the SM. In that case there is indeed a
nonzero neutrino-two-photon coupling, which arises at the one-loop
level and is negligibly small, of order $O(G_F^2)$ \cite{Dicus}.
Nevertheless, as pointed out in Ref. \cite{Dodelson}, the
introduction of massive neutrinos can enhance dramatically the
$\nu\bar\nu\gamma$ and $\nu\bar\nu\gamma\gamma$ vertices. While in
several SM extensions the neutrino-two-photon interaction is
proportional to the neutrino mass, there are some extensions, such
as left-right symmetric theories and the Zee model, where it is
proportional to a heavy Higgs scalar mass \cite{Liu}. This fact
may give rise to a significant enhancement of the $\nu \bar \nu
\gamma \gamma$ coupling, which can be parametrized in the
following way at the lowest dimension \cite{Nieves,Dodelson,Liu}

\begin{equation}
{\cal L}_{\bar \nu_i\nu_j \gamma \gamma}=\frac{1}{4 \,\Lambda^3}
\,\bar \nu_i\,\left(\alpha^{ij}_L P_L+\alpha^{ij}_R P_R \right)
\,\nu_j \tilde{F}^{\mu\nu} F_{\mu\nu} ,\label{Lnngg}
\end{equation}

\noindent where $\alpha^{ij}_{L,R}$ are dimensionless coupling
constants and $\Lambda$ is the new physics scale. Given this interaction, the
procedure described before can be used to obtain the
contribution to the decay $\decb$ from diagram 2(d) plus that in
which the photon pair emerges from the neutrino. We can write the
respective transition amplitude as follows

\begin{equation}
{\cal M}^{\alpha\beta\mu}=
\frac{i\,g}{4\,c_W\,\Lambda^3}\bar{\nu}_i(p_2) \varpi^\mu
\nu_j(p_1)\,\epsilon^{\lambda\rho\alpha\beta}\,{k_1}_\lambda
{k_2}_\rho, \label{MZnngg3}
\end{equation}

\noindent with

\begin{equation}
\varpi^\mu= \alpha_L^{ij}\,P_L\,
\left(\FMSlash{p}-\FMSlash{p}_1\right)^{-1}\,\gamma^\mu+
\alpha_R^{ij}\,P_R\,\gamma^\mu\,
\left(\FMSlash{p}-\FMSlash{p}_2\right)^{-1}.
\end{equation}

\noindent Again, we have neglected the neutrino masses since the
result is insensitive to them. The squared amplitude can be
written in a very short way:

\begin{equation}
|{\cal M}|^2=\frac{8 \,\alpha\,\pi \left(k_1\cdot k_2\right)^2}{3
c_W^2 s_W^2 m_Z^2\,\Lambda^6}\left(|\alpha^{ij}_L|^2F(p_1)
+|\alpha^{ij}_R|^2F(p_2)\right),
\end{equation}

\noindent with
\begin{equation}
F(p_1)=\frac{p_1 \cdot p_2}{\left(p-p_1\right)^2}\left(
m_Z^4-4\left(p\cdot p_1\right)^2-\frac{4\,m^2_Z\,p \cdot p_1}{p_1
\cdot p_2}\left(p-p_1\right)\cdot p_2\right).
\end{equation}

From the last expressions, the $\decb$ decay rate can be obtained
after Monte Carlo integration. Hereafter, we will consider the
contributions from the three SM neutrino species. The resulting
branching fraction is thus given by

\begin{equation}
{\mathrm BR}(\decb)=1.092\times 10^3\,
\sum_i\sum_j\left(|\alpha^{ij}_L|^2+|\alpha^{ij}_R|^2\right)\Bigg[\frac{1\,\mathrm
GeV}{\Lambda}\Bigg]^6,
\end{equation}

where the sums run over $\nu_e$, $\nu_\mu$, and $\nu_\tau$. After
using the experimental bound on $\decb$, we are left with

\begin{equation}
\Bigg[\frac{1\,\mathrm GeV}{\Lambda}\Bigg]^6
\sum_i\sum_j\left(|\alpha^{ij}_L|^2+|\alpha^{ij}_R|^2\right)\leq
2.85 \times 10^{-9} . \label{bound3}
\end{equation}

\noindent This bound is weaker than that obtained for a sterile
neutrino $\nu_s$ from the analysis of the Primakoff effect on the
process of $\nu_\mu N\to \nu_s N$ conversion in the external
Coulomb field of the nucleus $N$ \cite{Gninenko}.

\begin{figure}
\begin{center}
\epsfig{file=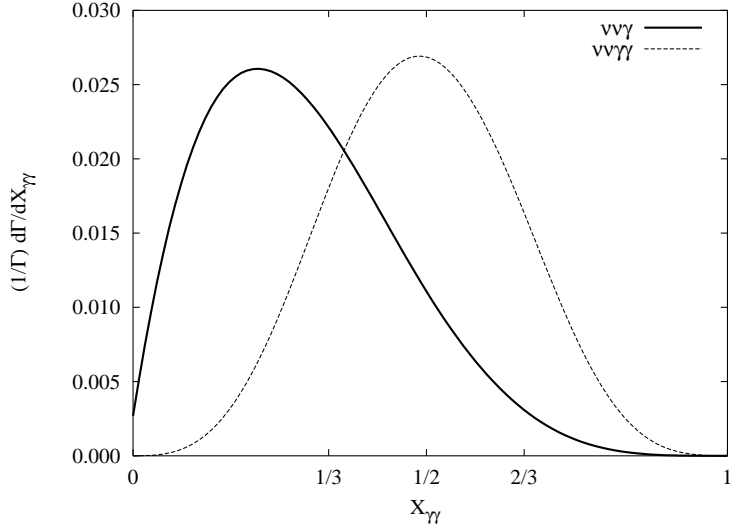,width=4in} \caption{Invariant mass
distribution of the photon pair
($X_{\gamma\gamma}=\sqrt{(k_1+k_2)^2}/m_Z$) in the rare decay
$\decb$. The contributions from the vertices $\nu\bar\nu\gamma$
and $\nu\bar\nu\gamma\gamma$ are shown separately.} \label{Xgg}
\end{center}
\end{figure}

At this point, we would like to note some interesting features of
the photon energy and invariant mass distributions of the decay
$\decb$. In Fig \ref{Xgg} we have plotted the distribution of the
invariant mass of the photon pair when the contribution from
either vertex $\nu\bar\nu\gamma$ or $\nu\bar\nu\gamma\gamma$ is
considered at a time. When only the $\nu\bar\nu\gamma$ interaction
contributes, the invariant mass peaks around
$X_{\gamma\gamma}=1/4$; on the other hand, when the
$\nu\bar\nu\gamma\gamma$ vertex alone contributes, the peak is
located around $X_{\gamma\gamma}=1/2$. A similar situation is
observed in Fig. \ref{Xgg}, where we have plotted the energy
distribution of the photon pair, and in Fig. \ref{Yg}, where it is
shown the energy distribution of an isolated photon. We can
observe that in both plots the peak of the curve accounting for
the $\nu\bar\nu\gamma$ contribution is shifted to the left with
respect to the curve resulting from the $\nu\bar\nu\gamma \gamma$
contribution. Therefore, in principle a proper set of cuts would
allow us to distinguish between the contributions from each
vertex. A more comprehensive analysis is however beyond the
present letter.

\begin{figure}
\begin{center}
\epsfig{file=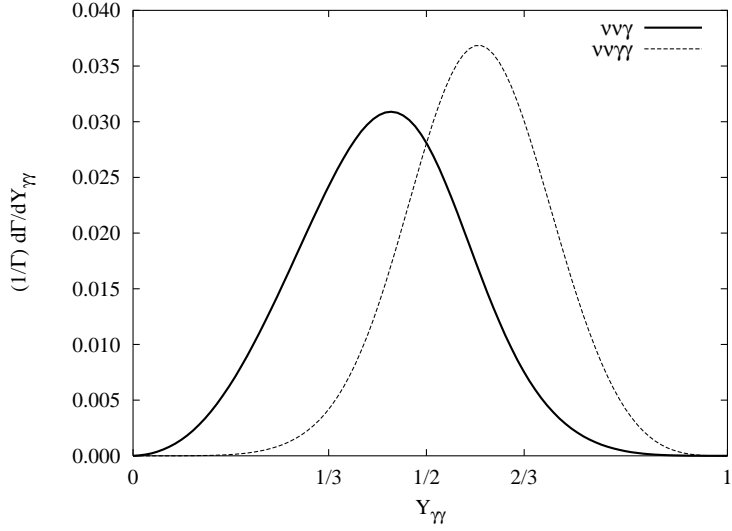,width=4in} \caption{Energy distribution
($Y_{\gamma\gamma}=(E_1+E_2)/m_Z$) of the photon pair in the rare
decay $\decb$. The contributions from the vertices
$\nu\bar\nu\gamma$ and $\nu\bar\nu\gamma\gamma$ are shown
separately.} \label{Ygg}
\end{center}
\end{figure}

\begin{figure}
\begin{center}
\epsfig{file=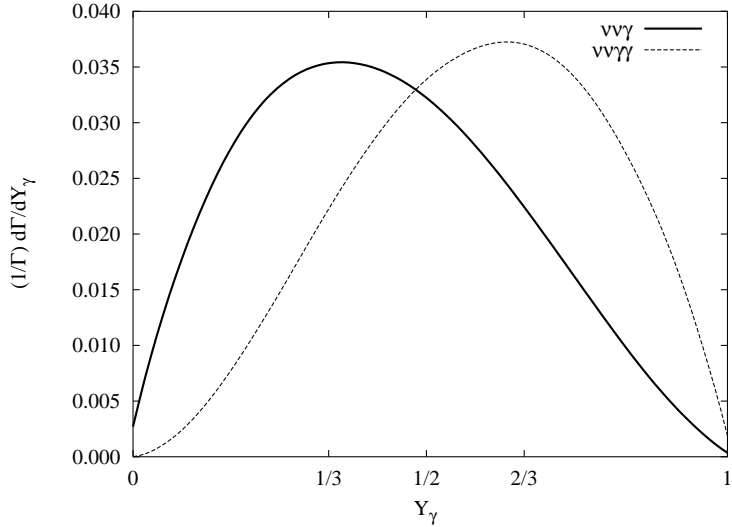,width=4in} \caption{Energy distribution
($Y_{\gamma}=2\,E_1/m_Z$) of a single photon in the rare decay
$\decb$. The contributions from the vertices $\nu\bar\nu\gamma$
and $\nu\bar\nu\gamma\gamma$ are shown separately.} \label{Yg}
\end{center}
\end{figure}

Finally, we will show that the decay $\decb$ can also be used to
bound the neutrino double radiative decay $\nu_j \to \nu_i \gamma
\gamma$. Neglecting the $\nu_i$ mass and after some calculation
one can obtain the following expression for the $\nu_j \to \nu_i
\gamma \gamma$ decay width

\begin{equation}
\Gamma_{\nu_j \to \nu_i \gamma
\gamma}=\frac{m^7_{\nu_j}}{2^{10}\,\Lambda^6\,\pi^3}
\left(|\alpha^{ij}_L|^2+|\alpha^{i j}_R|^2\right)
\int^1_0\int_x^1\left(1-x\right)\,x^2\,dy\,dx,
\end{equation}

\noindent which yields

\begin{equation}
\Gamma_{\nu_j\to\nu_i\gamma\gamma}=1.59\times
10^{-3}\left(|\alpha^{i j}_L|^2+|\alpha^{i j
}_R|^2\right)\Bigg[\frac{1\,\mathrm GeV}{\Lambda}\Bigg]^6
\Bigg[\frac{m_{\nu_j}}{1\,{\mathrm MeV}}\Bigg]^7 \;\,{\mathrm
s^{-1}}.
\end{equation}

\noindent The constraint of Eq. (\ref{bound3}) can then be
translated into a lower bound on the $\nu_j$ lifetime

\begin{equation}
\tau_{\nu_{j}}\ge  1.79\times10^{12}\Bigg[\frac{1\,\mathrm
MeV}{m_{\nu_j}}\Bigg]^7\;\, \mathrm{s},
\end{equation}

\noindent which is true provided that $m_{\nu_j}$ lies in the
range of a few tenths of a MeV, since in that case $\nu_j \to
\nu_i \gamma \gamma$ is the dominant decay channel \cite{Nieves}.
Our bound is one order of magnitude below than the one previously
found for the lifetime of a sterile neutrino \cite{Gninenko}.

In closing we emphasize that the rare decay $\deca$ gives rise to
a bound on the $\nu_\tau$ magnetic moment that is in excellent
agreement with other ones found recently. The bound obtained from
the rare decay $\decb$ is just one order of magnitude below. It
must be stressed that the current experimental bound on the
$\decb$ branching ratio is somewhat weak. In fact, it is of the
same order of magnitude than that on $\deca$. Some improvement is
expected from the data to be collected at a future linear
collider. One important feature of the decay $\decb$ is that it
can also be used to bound the neutrino-two-photon interaction. In
this respect, the resulting bound is weaker than that derived for
a sterile neutrino from the process of $\nu_\mu N\to\nu_s N$
conversion in the external Coulomb field of the nucleus $N$.
Finally, the bound on the neutrino-two-photon interaction also
allowed us to constrain the width of the decay
$\nu_{j}\to\nu_i\gamma\gamma$, which in turn can be translated
into a constrain on the $\nu_j$ lifetime as long as $m_{\nu_j}$
lies in the range of a few hundreds of KeV. The main advantage of
our procedure is that it is model-independent and relies on a few
assumptions.

\ack{We acknowledge support from CONACYT and SNI (M\'exico). One
of us (GTV) thanks A. Pukhov for valuable suggestions about the
implementation of the $\nu \bar\nu \gamma \gamma$ vertex into his
{\small CALCHEP} program.}

%


\begin{thebibliography}{99}

\bibitem{Fukuda}SK Collab., Y. Fukuda, {\it et al.}, Phys. Rev. Lett.
81 (1998) 1562; 86 (2001) 5651; 86 (2001) 5656.


\bibitem{Fukuda1}SNO Collab., Q. R. Ahmad {\it et al.},
Phys. Rev. Lett. 87 (2001) 071301.


\bibitem{Cisneros}A. Cisneros, Astrophys. Space Sci. 10 (1971) 87;
L. B. Okun, M. B. Voloshin, and M. I. Vysotky, Sov. Phys. JETP 64
(1986) 446.


\bibitem{Fujikawa}W. Marciano and A. Sirlin, Phys. Rev. D
22 (1980) 2695; W. Marciano and A. Sanda, Phys. Lett. B 67 (1977)
303; K. Fujikawa and R. Shrock, Phys. Rev. Lett. 45 (1980) 963; P.
Pal and L. Wolfenstein, Phys. Rev. D 25 (1981) 766.


\bibitem{Frank} See for instance M. Frank, Phys. Rev. D 60 (1999) 093005; Phys.
Lett. B 477 (2000) 208; M. A. B. Beg, W. J. Marciano, and M.
Ruderman, Phys. Rev. D 17 (1978) 1395; C.-K. Chua and W.-Y. P.
Hwang, Phys. Rev. D 60 (1999)073002.


\bibitem{Athana}LSND Collab., C. Athanassopoulos {\it et al.}, Phys. Rev. Lett.
81 (1998) 1774.


\bibitem{Pontecorvo}B. M. Pontecorvo, Zh. Eksp. Teor. Fiz. 36 (1959) 1615; H. Y.
Chiu and P. Morrison, Phys. Rev. Lett. 5 (1960) 573.

\bibitem{Levine} M. J. Levine, Nuovo. Cim. A 48 (1967) 67.

\bibitem{Nieves}J. F. Nieves, Phys. Rev.  D 28 (1983) 1664; R. K. Ghosh, Phys.
Rev. D 29 (1984) 493.


\bibitem{Acciari}L3 Collab., M. Acciari {\it et al.}, Phys. Lett. B 412 (1997) 201.


\bibitem{Maya}M. Maya, M. A. P\'erez, G. Tavares-Velasco, and B. Vega, Phys. Lett.
B 434 (1998) 354.


\bibitem{Maltoni} M. Maltoni and M. I. Vytsosky, Yad. Fiz. 62
(1999) 1278 ; Phys. Atom. Nucl. 62 (1999) 1203.


\bibitem{Hernandez}J. M. H\'ernandez, M. A. P\'erez, G. Tavares-Velasco and J. J.
Toscano, Phys. Rev. D 60 (1999) 013004.


\bibitem{Larios} F. Larios, M. A. P\'erez, G. Tavares-Velasco and J. J. Toscano,
Phys. Rev. D 63 (2001) 113014.


\bibitem{OPAL}L3 Collab., O. Adriani {\it et al.}, Phys. Lett. B 295 (1992) 337;
OPAL Collab., P. Acton {\it et al.}, Phys. Lett. B 311 (1993) 391.


\bibitem{Grotch} H. Grotch and R. Robinett, Z. Phys. C 39 (1988)
553; T. M. Gould and I. Z. Rhotstein, Phys. Lett. B 333 (1994)
545.


\bibitem{Escribano} R. Escribano and E. Masso, Phys. Lett. B 395 (1997) 369.


\bibitem{Cooper-Sarkar}A. M. Cooper-Sarkar {\it et al.}, Phys.
Lett. B 280 (1992) 153.


\bibitem{Ayala}A. Ayala, J. C. D'Olivo, and M. Torres, Phys. Rev.
D 59 (1999) 111901; R. Barbieri and R. N. Mohapathra, Phys. Rev.
Lett. 61 (1988) 27.


\bibitem{Krakauer} D. A. Krakauer {\it et al.}, Phys. Lett. B 252 (1990) 177;
R. C. Allen {\it et al.}, Phys. Rev. D 47 (1993) 11.


\bibitem{VEGAS}G. P. Lepage, J. Comput. Phys. 27 (1978) 192.


\bibitem{MonteCarlo}F. James, Monte Carlo Phase Space, CERN 68-15
(1968).

\bibitem{COMPHEP} A. Pukhov {\it et al.}, {\it COMPHEP, a
package for evaluation of Feynman Diagrams and integration over
multi-particle phase space}, preprint hep-ph/9908288.


\bibitem{Martinez} See for instance F. Larios, R. Mart\'{\i}nez, and M. A. P\'erez,
Phys. Lett. B 345 (1995) 259; K. S. Babu, T. M. Gould, and I. Z.
Rhotstein, Phys. Lett. B 321 (1994) 140.


\bibitem{GellMann}M. Gell-Mann, Phys. Rev. Lett. 6 (1961) 70.

\bibitem{Dicus}D. A. Dicus and W. A. Repko, Phys. Rev. D 48 (1993)
5106; Phys. Rev. Lett. 79 (1997) 569; A. Abbasabadi, A. Devoto, D.
A. Dicus, and W. Repko, Phys. Rev. D 59 (1999) 013012.


\bibitem{Dodelson} S. Dodelson and G. Feinberg, Phys. Rev. D 43 (1991) 913.


\bibitem{Liu}J. Liu, Phys. Rev. D 44 (1991) 2879;


\bibitem{Gninenko} S. N. Gninenko and N. V. Krasnikov, Phys. Lett. B 450 (1999)
165.


\end{thebibliography}
\end{document}